# Biaxial Tensile Strain Enhances Electron Mobility of Monolayer Transition Metal Dichalcogenides

Jerry A. Yang,[1] Robert K. A. Bennett,[1] Lauren Hoang,[1] Zhepeng Zhang,[2] Kamila J. Thompson,[1] Andrew J. Mannix[2,3], and Eric Pop[1,2,4,*]

[1]*Department of Electrical Engineering, Stanford University, Stanford, CA 94305, USA*
[2]*Department of Materials Science & Engineering, Stanford University, Stanford, CA 94305, USA*
[3]*Stanford Institute for Materials and Energy Sciences, SLAC National Accelerator Laboratory, Menlo Park, CA 94305, USA*
[4]*Precourt Institute for Energy, Stanford University, Stanford, CA 94305, USA*

**ABSTRACT:** Strain engineering can modulate the material properties of two-dimensional (2D) semiconductors for electronic and optoelectronic applications. Recent theory and experiments have found that uniaxial tensile strain can improve the electron mobility of monolayer $MoS_2$, a 2D semiconductor, but the effects of biaxial strain on charge transport are not well-understood in 2D semiconductors. Here, we use biaxial tensile strain on flexible substrates to probe the electron mobility in monolayer $WS_2$ and $MoS_2$ transistors. This approach experimentally achieves ~2× higher on-state current and mobility with ~0.3% applied biaxial strain in $WS_2$, the highest mobility improvement at the lowest strain reported to date. We also examine the mechanisms behind this improvement through density functional theory simulations, concluding that the enhancement is primarily due to reduced intervalley electron-phonon scattering. These results underscore the role of strain engineering 2D semiconductors for flexible electronics, sensors, integrated circuits, and other opto-electronic applications.

Two-dimensional (2D) semiconductors have gained significant interest for electronic and optoelectronic devices due to their atomically thin structure,[1] theoretically pristine van der Waals interfaces,[1,2] and potential for heterogeneous integration into monolithic three-dimensional (3D) computing systems.[1,3] Tungsten disulfide, $WS_2$, and molybdenum disulfide, $MoS_2$, are commonly explored 2D transition metal dichalcogenide (TMD) semiconductor because they can be grown in single layers with good control,[4,5] have high electron mobilities at atomic-scale thicknesses,[6] and may exhibit electronic ambipolarity as a single or multi-layer material.[7,8] However, significant challenges remain in incorpo-rating WS$_2$ and other TMDs into large-scale heterogeneous integration schemes, one of which is improving the electrical performance metrics, such as the mobility and on-state current of TMD-based field-effect transistors (FETs). While techniques including channel doping,[9] interface engineering,[10,11] and contact engineering[11] have been used to improve electrical performance of WS$_2$ transistors, the potential for strain engineering to enhance device performance remains to be explored.

Recently, an experimental study has shown that the mobility of monolayer MoS$_2$ is nearly doubled with less than 1% tensile strain,[12] while theoretical efforts have predicted effects of both tensile and compressive strain on the band structure[13,14] and mobility[15–17] of various TMDs. In particular, simulations show that biaxial strain should have a stronger effect on monolayer MoS$_2$ mobility compared to uniaxial strain.[18] The study of strain on transistor performance also has industrial relevance; for example, tensile and compressive strain are built into every modern silicon transistor to enhance the electron and hole mobilities, respectively.[19] However, despite significant efforts in understanding strain effects in optical emission,[20] the effect of strain on the electronic properties of WS$_2$ and MoS$_2$, particularly under non-uniform and non-uniaxial strain fields, remains to be explored. In addition, theoretical studies have also predicted that WS$_2$ should have the highest intrinsic mobility and saturation velocity among the basic TMDs.[15,21]

In this work, we study the effect of biaxial tensile strain on the optical properties and electron mobility of alumina-encapsulated monolayer WS$_2$ and non-encapsulated monolayer MoS$_2$ transistors fabricated on bendable substrates. We find that the electron mobility and on-state current increase by ~100% at ~0.3% biaxial tensile strain for WS$_2$ and ~60% at the same strain for MoS$_2$, and the encapsulation significantly alters the optical band gap and phonon modes of WS$_2$ under tensile strain compared to non-encapsulated WS$_2$ films. We correlate our experimental results to density functional theory simulations for WS$_2$, confirming that the mobility and on-state current enhancements result from reduced intervalley scattering but not changes in electron effective mass. These results highlight strain engineering as a key mobility booster for improving the performance of TMD-based electronics.

**RESULTS AND DISCUSSION**

**Device Structure and Biaxial Strain Tool Operation. Figure 1** shows the device structure of our TMD field-effect transistors fabricated on free-standing, flexible, and transparent substrates of polyethylene naphthalate (PEN, 125 μm thick). Monolayer WS$_2$ and monolayer MoS$_2$ were grown on separate substrates, then transferred onto the PEN film, which had Al$_2$O$_3$ gate dielectrics on pre-patterned metal back-gates (**Figure 1a**). Ni/Au source and drain contacts were patterned on top by optical lithography and lift-off, and WS$_2$ devices were encapsulated by a thin layer of AlO$_x$. **Figure 1b** shows the top-down image of one finished device and **Figure 1c** shows the flexible substrate with several hundred devices. Additional details regarding TMD growth and layer transfer are presented in the Methods section, the full fabrication process is discussed in Supporting Information Section 1, and the effect of the sub-stoichiometric AlO$_x$ on the electrical performance of WS$_2$ transistors is presented in Supporting Information Section 2.

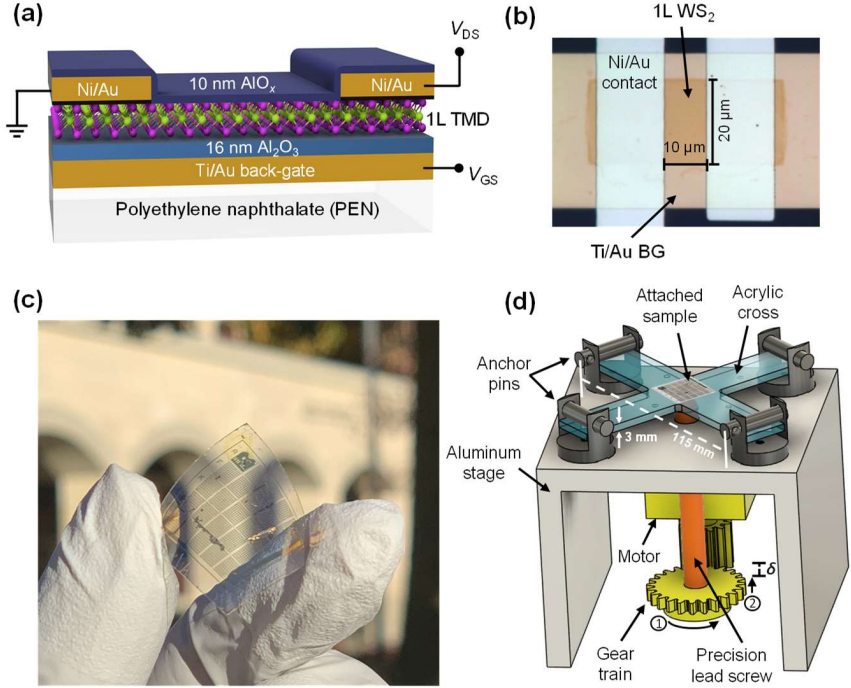

**Figure 1.** TMD device schematic and strain tool. (a) Schematic of back-gated (BG) monolayer (1L) WS$_2$ transistor on a polyethylene naphthalate (PEN) flexible substrate, with source and drain contacts (~5 nm Ni capped with ~40 nm Au), ~16 nm thick Al$_2$O$_3$ back-gate dielectric, and ~10 nm thick AlO$_x$ encapsulation. (b) Top-view optical image of 1L WS$_2$ transistor with channel width $W \approx 20$ μm length $L \approx 10$ μm. (c) Camera photo of flexed substrate chiplet with several hundred WS$_2$ devices. (d) Biaxial strain tool schematic. A motor drives a gear train, which turns a precision lead screw, deflecting the center of an acrylic cross to which the sample is attached. The deflection imparts biaxial strain into the PEN sample.

We applied strain to our samples with a motorized stage shown in **Figure 1d.** (Also see Supporting Information Figure S2 for additional details.) An external Arduino microcontroller operates a precision lead screw that deflects an acrylic cross, to which the PEN sample is attached. The strain at the center of the cross can then be estimated as $\varepsilon \approx 4t\delta/D^2$, where $t = 3$ mm is the thickness of the acrylic cross, $\delta$ is the distance of deflection in millimeters, and $D = 115$ mm is the distance between two non-adjacent anchor pins. Previous modeling work has shown that this strain geometry produces a biaxial, radially-symmetric tensile strain field within 3 mm of the center, with the strain decaying to about 75% at 3 mm away from the center.[22] Due to potential strain fields induced in the WS$_2$ from device fabrication and encapsulation, all strain measurements are reported relative to the as-fabricated devices after attachment to the acrylic cross and before loading onto the strain tool.

**Spectroscopic Measurements of WS$_2$ Devices Under Strain.** We use Raman and photoluminescence (PL) spectroscopy to confirm strain transfer from the acrylic cross to the WS$_2$ devices (**Figure 2**). We monitor the positions of the A$_1$' peak at ~418 cm$^{-1}$, E' peak at ~356 cm$^{-1}$, 2LA(M) peak at ~348 cm$^{-1}$, A$^0$ exciton peak at ~1.95 eV, and the A$^0$ exciton shoulder peak at ~1.92 eV (usually assigned to the A$^-$ trion peak signature[20,23]), respectively, as these peaks have been shown to be sensitive to strain.[20] We measure the spectra at 9 points within the device channel for each strain level, then use linear regression to calculate the peak shift rate under strain for each peak. We note that the AlO$_x$ encapsulation alters the optical signatures of WS$_2$ (compared to unencapsulated samples), consistent with previous findings;[20,24,25] however, this does not affect our conclusions because all our strain measurements are relative to the strain tool itself. In other words, optical signatures of unencapsulated WS$_2$ are not used to calibrate the encapsulated samples.

**Figure 2a** shows the Raman spectra of a WS$_2$ device before encapsulation, after AlO$_x$ encapsulation but before imparting strain, and after strain. Qualitatively, the AlO$_x$ broadens the 2LA(M) peak and increases the prominence of the E' peak in the Raman spectrum, whereas the A$_1$' peak intensity remains unaffected. After encapsulation, the 2LA(M) and A$_1$' peaks redshift, while the E' peak slightly blueshifts. **Figure 2b** showcases boxplots for the Raman peak positions at each strain level and the extracted strain rate. Under strain, the 2LA(M) peak redshifts by 13 ± 5.0 cm$^{-1}$/% strain, and the E' peak blueshifts by 1.9 ± 1.1 cm$^{-1}$/% strain, whereas the A$_1$' peak does not appear to shift. The E' and A$_1$' peak shift rates in our encapsulated devices differ from previously reported strain rates of unencapsulated monolayer WS$_2$ under biaxial tensile strain,[20] likely because the AlO$_x$ passivation step removes surface oxygen and water adsorbates,[26] inducing partial electron doping. (This is also seen as a negative threshold voltage shift in Supporting Information Figure S3.) A reference sample from the same WS$_2$ growth and fabrication run was used to measure the optical strain response of unencapsulated devices, finding it consistent with previous literature (Supporting Information Figure S4).

**Figure 2c** shows the PL spectra of the same WS$_2$ device as **Figure 2a** before encapsulation, after AlO$_x$ encapsulation but before imparting strain, and after strain. The A$^0$ exciton peak redshifts and becomes more symmetric (as indicated by the suppressed left shoulder peak) after AlO$_x$ encapsulation and after imparting strain. The combined effect of encapsulation and strain causes a ~100 meV redshift of the A$^0$ exciton peak. **Figure 2d** shows how the A$^0$ exciton and shoulder peak positions evolve with strain. Consistent with our density functional theory simulations (shown in Supporting Information

Section 8 and discussed later), the $A^0$ exciton peak redshifts by ~120 ± 61 meV/% strain. This is also consistent with previous reports of WS$_2$ under biaxial tensile strain with a similar strain setup.[20]

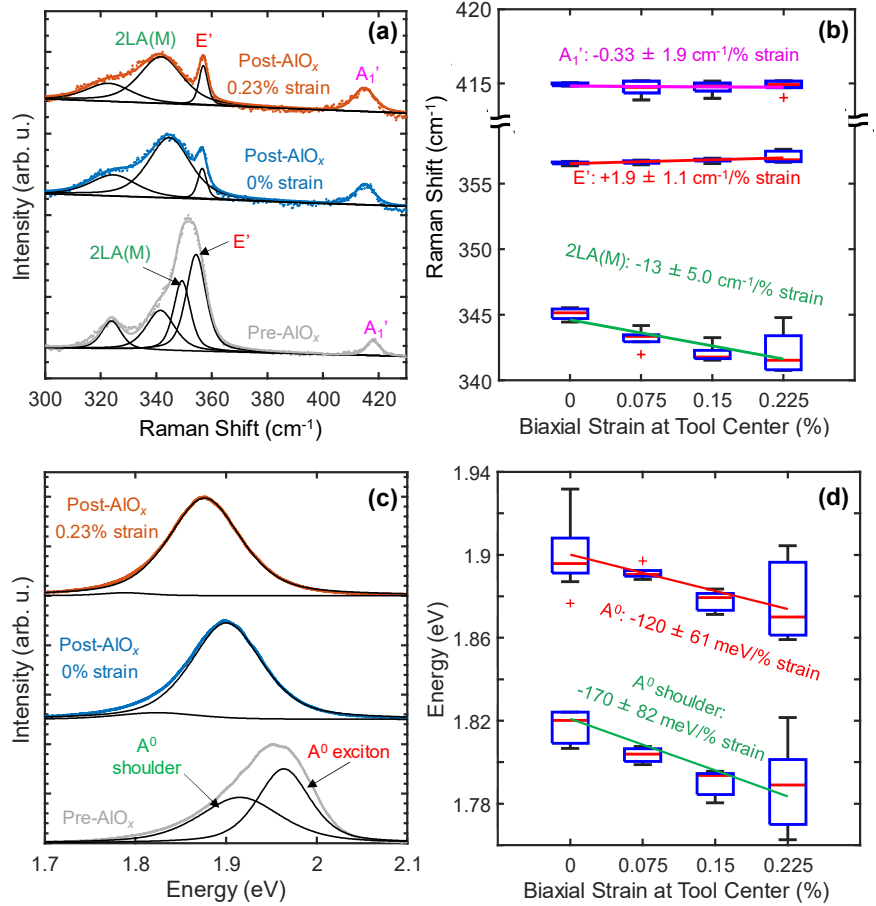

**Figure 2.** Raman and photoluminescence (PL) response of monolayer WS$_2$ to biaxial strain. (a) Raman spectra of a device with similar geometry as **Figure 1b** before encapsulation (gray), after AlO$_x$ encapsulation (blue), and after AlO$_x$ encapsulation with biaxial strain (orange). The data (symbols) are fitted with a superposition of Gaussian and Lorentzian peaks (black curves). (b) Extracted phonon shift rates of the A$_1$' (pink), E' (red), and 2LA(M) (green) Raman peaks of the AlO$_x$-encapsulated WS$_2$ as a function of biaxial strain applied, with box plots indicating the median (red line), first, and third quartiles (blue box) across nine measured locations in the device channel. (c) PL spectra of the device at the same locations in the channel, under the same conditions and color labeling as in (a). (d) Extracted exciton shift rates of the $A^0$ exciton (red) and its left shoulder peak (green), with similar notation and same number of points as in (b).

**Electrical Measurements of WS$_2$ Devices Under Strain.** We next perform electrical measurements of our devices with applied biaxial strain. Our setup enables direct probing of our transistors under strain in a probe station, in air ambient (see Supporting Information Figure S2). We measure multiple devices on two separate flexible substrates to exclude possible degradation from multiple strain cycles. **Figure 3a** displays the drain current ($I_D$) vs. gate voltage overdrive ($V_{GS} - V_T$) of the device from **Figure 1b** at $V_{DS}$ = 1 V. Here $V_T$ is the threshold voltage estimated by the constant-current method[27] with a threshold current of $I_D = 10$ nA/μm, consistent with the International Roadmap for Devices and Systems (IRDS) 2022 threshold for high-performance devices.[28] (We compare the constant-current,[27] linear extrapolation,[27] and Y-function[29] methods of $V_T$ extraction for our devices in Supporting Information Section 4.)

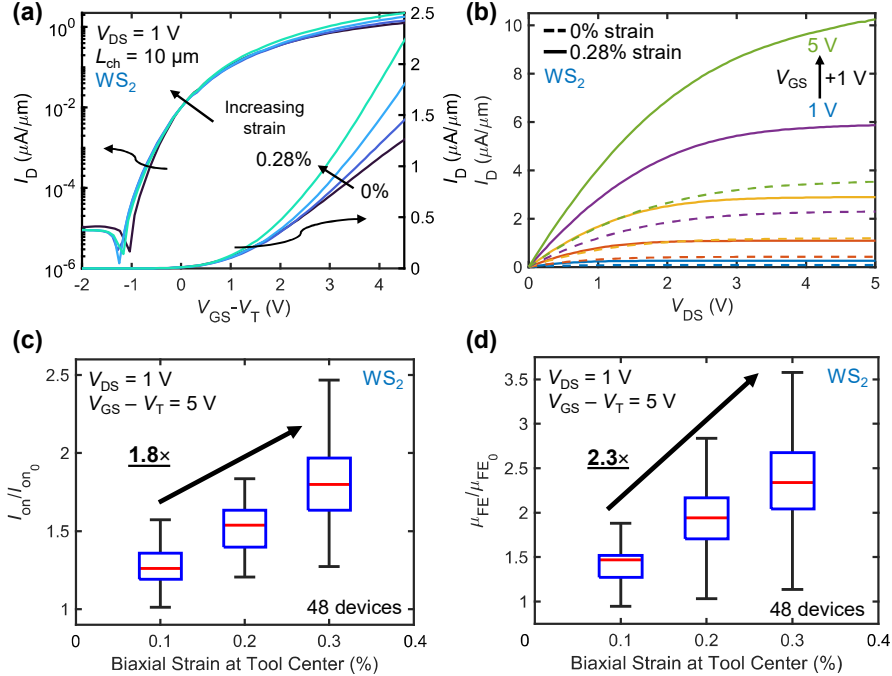

**Figure 3.** Strained WS$_2$ device measurements. (a) Transfer characteristics ($I_D$ vs. $V_{GS}$-$V_T$) of the device from **Figure 1b** ($W = 20$ μm, $L_{ch} = 10$ μm) at different levels of applied biaxial tensile strain, where $V_T$ is the constant-current threshold voltage. Same data are plotted both on a log scale (left y-axis) and a linear scale (right y-axis). (b) Output characteristics ($I_D$ vs. $V_{DS}$) of the same device at 0% (dashed) and 0.28% (solid) applied biaxial tensile strain for $V_{GS} = 1$ V (blue), 2 V (red), 3 V (yellow), 4 V (purple), and 5 V (green). Significant saturation current enhancement is seen with strain applied. (c) On-current ($I_{on}$) normalized to the initial unstrained values ($I_{on,0}$) for 48 transistors as a function of applied biaxial strain, with the box plots showing the median across devices (red lines), first and third quartiles (blue box), and maximum and minimum (top and bottom horizontal lines, respectively). (d) Field-effect mobility ($\mu_{FE}$) normalized to the initial unstrained values for the same 48 transistors as in (c) as a function of applied strain. $I_{on}$ and $\mu_{FE}$ values were extracted at $V_{DS} = 1$ V and $V_{GS}$ - $V_T = 4$ V.

In **Figure 3a** we plot $I_D$ vs. gate overdrive voltage to account for possible shifts in $V_T$ and electron density due to strain, for several values of biaxial strain up to 0.28% (estimated based on its location relative to the center of the sample). **Figure 3b** shows the corresponding $I_D$ vs. $V_{DS}$ measurements at 0% (dashed lines) and 0.28% (solid lines) biaxial tensile strain for $V_{GS}$ from 1 V to 5 V. The drain current increases with strain: for $V_{GS} = V_{DS} = 5$ V, the drain current nearly triples from ~3.5 μA/μm to ~10.2 μA/μm at 0.28% strain. We note that the output characteristics display current saturation with a nearly quadratic (in $V_{GS} - V_T$) dependence, emphasizing this is a classic, "textbook-like" long-channel field-effect transistor. Additional information about the (small) hysteresis and gate leakage current for this device can be found in Supporting Information Section 5.

We estimate the on-state current ($I_{on}$) and field-effect mobility ($\mu_{FE}$) from measurements (*e.g.* **Figure 3a**) in the linear region at the same gate overdrive ($V_{GS} - V_T = 5$ V). In the linear regime, when $V_{DS} < V_{GS} - V_T$, the field-effect mobility is $\mu_{FE} = (\partial I_D/\partial V_{GS})L_{ch}/(W_{ch}C_{ox}V_{DS})$, where $C_{ox} \approx 378$ nF/cm$^2$ is the gate capacitance per unit area. (The Al$_2$O$_3$ thickness was measured by ellipsometry on a reference silicon piece, and its dielectric constant was measured with metal-insulator-metal capacitors.[12]) The transistor shown in **Figure 3a** has $I_{on} \approx 1.8$ µA/µm and $\mu_{FE} \approx 12$ cm$^2$ V$^{-1}$ s$^{-1}$ without applied strain. At 0.3% applied biaxial tensile strain, the transistor has $I_{on} \approx 3.2$ µA/µm and $\mu_{FE} \approx 30.5$ cm$^2$ V$^{-1}$ s$^{-1}$. Therefore, we achieve 125% improvement in $\mu_{FE}$ at ~0.28% biaxial tensile strain for this device. [The current enhancement appears greater in saturation for the same device in **Figure 3b**, which does not adjust for $V_T$, because $V_T$ decreases with strain. This effect is magnified by $\propto (V_{GS} - V_T)^2$ in saturation. See Supporting Information Figure S5 for changes in $V_T$ with strain.]

We measure 47 other transistors with channel lengths from $L_{ch} = 2$ µm to 15 µm on the same substrate and perform electrical measurements with applied strain. To account for device variation, we extract the $I_{on}$ and $\mu_{FE}$ for all devices at all strain levels and report the improvements in $I_{on}$ and $\mu_{FE}$ relative to their initial values without applied strain, as summarized in **Figures 3c** and **3d.** On average, for WS$_2$, $I_{on}$ increases by a factor of 1.8 ± 0.4, and $\mu_{FE}$ increases by a factor of 2.3 ± 0.5 at 0.3% applied tensile strain at tool center, compared to the initial values without strain. If $V_T$ is estimated with the linear extrapolation and Y-function methods, the improvement factor is higher for both $I_{on}$ and $\mu_{FE}$ (see Supporting Information Figure S5). Our $\mu_{FE}$ improvements due to biaxial tensile strain exceed previous improvements seen in uniaxial tensile strain studies on MoS$_2$ field-effect transistors.[12]

In **Figure 3**, we report device behavior up to 0.3% applied biaxial tensile strain. At 0.4% strain at tool center, over half of the devices showed a degradation in $I_{on}$ and $\mu_{FE}$ or did not generate enough current for constant-current $V_T$ extraction, likely due to strain-induced defects in the gate dielectric or poorer adhesion at the contact-WS$_2$ interface. (See Supporting Information Section 6 for device relaxation and degradation after higher strain.) Due to the nonuniform strain field imparted by the tool geometry,[22] we examine the impact of device location on $I_{on}$ and $\mu_{FE}$ improvement. Supporting Information Section 7 shows heat maps for the strain evolution of $I_{on}$ and $\mu_{FE}$ for each measured device. There was no significant difference between the devices that experienced more longitudinal (*i.e.*, along-channel) strain compared to the devices that experienced more transverse (*i.e.*, across-channel) strain. We attribute this to the isotropic in-plane conduction band valleys in $WS_2$[21] and random orientation of grain boundaries in the channels. Therefore, tensile strain in any direction may have an additive effect of increased $I_{on}$ and $\mu_{FE}$, producing larger $I_{on}$ and $\mu_{FE}$ improvements for biaxial tensile strain compared to uniaxial tensile strain.

**Electrical Measurements of MoS₂ Devices Under Strain.** We repeat the strained electrical measurements for one flexible chiplet of $MoS_2$ transistors with the same device structure and fabrication process, but without the $AlO_x$ capping layer, which can introduce excessive doping and hysteresis in $MoS_2$ unless annealed at temperatures beyond the thermal budget of the PEN.[30] **Figures 4a and 4b** display the $I_D$ vs. $V_{GS} - V_T$ and the corresponding $I_D$ vs. $V_{DS}$ curves respectively for a device with channel width $W = 20$ μm and channel length $L_{ch} = 12$ μm as a function of strain, up to 0.23% strain. This device showed an increase in $I_{on}$ from ~0.44 μA/μm to ~0.63 μA/μm and an increase in $\mu_{FE}$ from ~5.4 cm² V⁻¹ s⁻¹ to ~9.1 cm² V⁻¹ s⁻¹ at ~0.23% strain, equating to a 40% increase in $I_{on}$ and 70% increase in $\mu_{FE}$ at $V_{DS} = 1$ V and $V_{GS} - V_T = 4$ V. Similar to $WS_2$, the saturation current for this $MoS_2$ nearly triples from 1.1 μA/μm to 3.1 μA/μm with 0.23% biaxial tensile strain at $V_{GS} = V_{DS} = 5$ V.

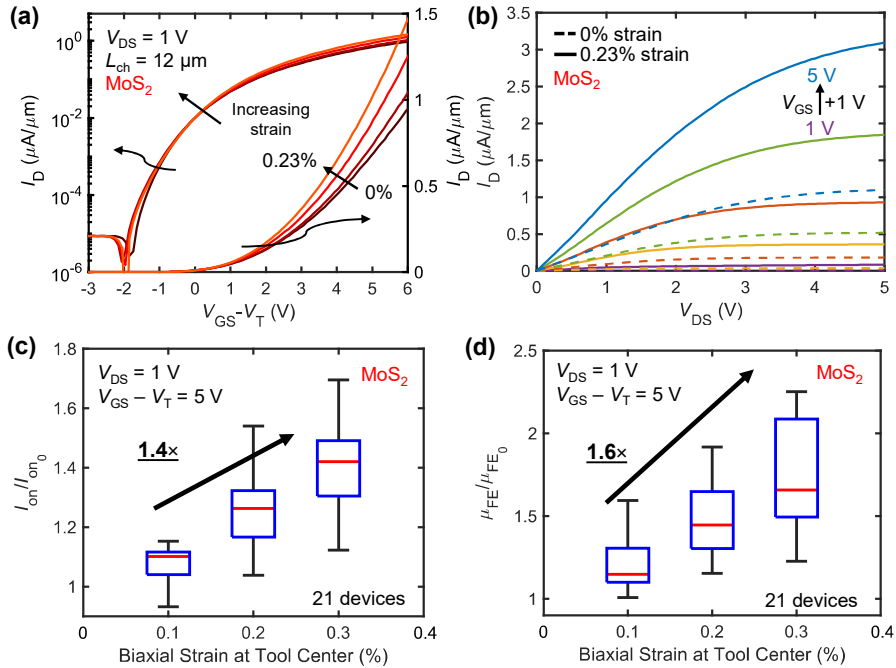

**Figure 4.** Strained MoS₂ device measurements. (a) Transfer characteristics ($I_D$ vs. $V_{GS}$-$V_T$) of a device with width $W = 20$ μm and channel length $L_{ch} = 12$ μm at different levels of applied biaxial tensile strain, as in **Figure 3a**. (b) Output characteristics ($I_D$ vs. $V_{DS}$) of the same device at 0% (dashed) and 0.23% (solid) applied biaxial tensile strain for $V_{GS} = 1$ V (purple) to 5 V (blue). Significant saturation current enhancement is seen with strain applied, as in the WS₂ device. (c) On-current ($I_{on}$) normalized to the initial unstrained values ($I_{on,0}$) for 21 transistors as a function of applied biaxial strain, as in **Figure 3c**. (d) Field-effect mobility ($\mu_{FE}$) normalized to the initial unstrained values ($\mu_{FE,0}$) for the same 21 transistors as in (c) as a function of applied strain. $I_{on}$ and $\mu_{FE}$ values were extracted at $V_{DS} = 1$ V and $V_{GS} - V_T = 5$ V.

We measure 20 other transistors with channel lengths from $L_{ch} = 2$ μm to 15 μm on the same substrate, extracting and reporting the $I_{on}$ and $\mu_{FE}$ in a similar manner as for the WS$_2$ sample. On average, for MoS$_2$, $I_{on}$ increases by a factor of 1.4 ± 0.07, and $\mu_{FE}$ increases similarly by a factor of 1.5 ± 0.16 at 0.3% applied biaxial tensile strain at tool center, compared to the initial values without strain. These values are slightly lower than those of WS$_2$ because the measured devices were on average 5% farther away from the center of the sample than those of the WS$_2$ sample. (See Supporting Information Section 7 for a discussion of the strain fields induced by the tool.)

**Simulations and Discussion.** To understand the causes of mobility enhancement in our biaxially-strained devices, we turn to density functional theory (DFT) simulations with spin-orbit coupling, which plays a stronger role in W-based materials.[31] From previous work on TMDs and other semiconductors (including Si), one can expect that strain affects the energy band gap, the electron effective masses, and electron-phonon scattering rates.[12,13,15,18,32,33]

From our experiments (**Figure 2d**) and simulations (Supporting Information Figure S12), biaxial strain reduces the WS$_2$ band gap, and consequently the transistor threshold voltage $V_T$ (Supporting Information Figure S5). However, the $V_T$ shift due to strain is small (< 0.5 V for our devices, depending on extraction method) compared to the gate voltage overdrive used (4 to 5 V). Because we report $I_{on}$ and $\mu_{FE}$ at a constant gate overdrive, this accounts for the $V_T$ shift due to strain, ruling out band gap reduction as a primary contributor to strain-related improvements of our devices. Regarding changes of effective mass, both our DFT simulations (**Figure 5a**) and previous work[15] suggest that for the range of biaxial strains applied in our experiments, the curvature of the conduction band at the K point changes only minimally and cannot be responsible for the observed mobility enhancement.

Our DFT simulations show a much greater effect of biaxial strain on the K-Q valley separation ($\Delta E_{QK}$), which is known to control electron-phonon intervalley scattering and therefore mobility.[33] (The Q point resides on the T line approximately halfway between the K and the Γ points.[34]) **Figure 5b** plots the evolution of $\Delta E_{QK}$ with both uniaxial and biaxial tensile strain, showing a nearly linear increase within the range considered here. Uniaxial tensile strain increases $\Delta E_{QK}$ by 86 ± 1.9 meV/% strain, and biaxial tensile strain increases $\Delta E_{QK}$ by almost twice as much, 169 ± 2.1 meV/% strain. The greater change in $\Delta E_{QK}$ with biaxial tensile strain suggests a higher mobility improvement for WS$_2$ under these conditions. We note that due to the layout of our test chiplets, our devices likely experienced unequal tensile strains in multiple directions within the applied biaxial strain field (Supporting Information Figure S11). This signifies that it may be possible to achieve the same mobility enhancement in WS$_2$ without precisely tuning the directionality or principal strain axes.

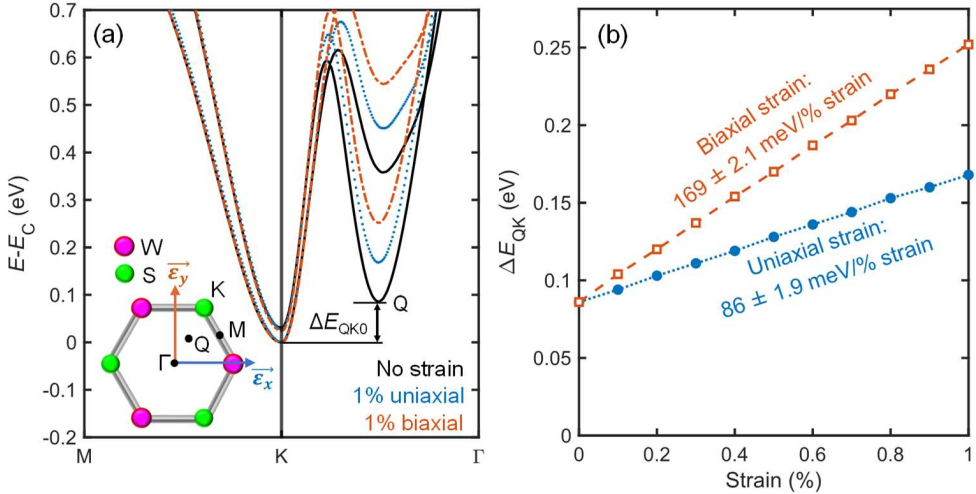

**Figure 5.** Density functional theory (DFT) calculations. (a) Simulated conduction band energy of WS$_2$ without strain (black solid line), 1% uniaxial strain (blue dotted line), and 1% biaxial strain (orange dashed line), relative to the conduction band edge. Two lowest conduction bands are seen for each case, due to spin splitting. Inset shows the unit cell and applied strain vectors $\vec{\varepsilon_x}$ and $\vec{\varepsilon_y}$ used in the simulation. (b) Energy separation between the K and Q conduction band valleys ($\Delta E_{QK}$) as a function of applied tensile strain. $\Delta E_{QK}$ increases with strain, and at a higher rate for biaxial tensile strain than for uniaxial tensile strain, indicating that strain causes a reduction of K-Q intervalley scattering. The calculated $\Delta E_{QK0}$ without strain (~87 meV) accounting for spin-orbit coupling is consistent with the report of Gaddemane et al.[31]

Finally, we comment on the possible role of tensile strain on defects and contacts. Defect trap states in the TMD or at the TMD-gate dielectric interface may facilitate hopping-like electron transport. Applied tensile strain can modulate the energy locations of these trap states in relation to the conduction band edge, increasing carrier lifetime[35] and improving channel conduction. This process is not well-understood, and future work ought to explore the fundamental role of strain on defects in TMDs. Tensile strain may also lower Schottky barriers[36] at TMD transistor contacts, reducing the contact resistance.[37,38] While these effects have been shown in MoS$_2$, our TMD device dimensions are large (~10 μm), reducing the influence of the contacts. We do not see a dependence of the strain enhancement effects on channel length (between 2 and 15 μm), as shown in Supporting Information Figure S13, suggesting that the contacts do not play a dominant role in our measurements.

## CONCLUSIONS

We have investigated the on-state current and mobility enhancement of CVD-grown monolayer WS$_2$ and MoS$_2$ transistors with applied biaxial tensile strain. We achieved this by straining devices fabricated on flexible PEN substrates, through the deflection of an acrylic cross with a motorized precision lead screw. On average, at 0.3% biaxial tensile strain, the WS$_2$ devices experienced a 100% increase of electron mobility and on-state drain current, the largest mobility improvement of TMD transistors with the lowest applied tensile strain to date. Based on DFT simulations, we attribute this improvement to reduced intervalley scattering from the increased energy separation $\Delta E_{QK}$ under tensile strain. Our results highlight strain engineering as a key future performance enhancer for TMD devices in both flexible and traditional rigid electronics and opto-electronics.

## METHODS

**Device Fabrication and Layer Transfer.** We fabricated locally back-gated monolayer TMD field-effect transistors on flexible, transparent polyethylene naphthalate (PEN, Dupont Teijin Films, 125 μm thick) sheets. Monolayer WS$_2$ was grown by chemical vapor deposition (CVD) onto *c*-plane sapphire substrates dip-coated in aqueous ammonium metatungstate (AMT) precursor (see Supporting Information Section 1). Monolayer MoS$_2$ was grown by solid-source CVD onto 90 nm SiO$_2$/Si substrates.[39] Separately, PEN substrates were prepared with Al$_2$O$_3$ dielectrics on pre-patterned metal back-gates (**Figure 1a**). After growth, the TMDs were then transferred with a PMMA/polystyrene bilayer stamp to the PEN patterned the local back-gates and Al$_2$O$_3$ gate oxide. The 5 nm/ 45 nm Ni/Au contacts are patterned by optical lithography and lift-off, and additional steps are provided in Supporting Information Section 1. To improve the device performance, we encapsulated the devices with sub-stoichiometric AlO$_x$ deposited by first electron-beam evaporating a ~1.5 nm Al seed layer followed by ~8.5 nm Al$_2$O$_3$ *via* thermal atomic layer deposition at 130 °C. The sub-stoichiometric AlO$_x$ encapsulation anneals the sample and imparts some strain into the WS$_2$ due to thermal expansion effects.[30,40] Due to the low glass transition temperature of PEN, we carefully tuned the thermal budget to ensure no degradation in the substrate due to processing. Once device fabrication was complete, the sample was attached to the 3 mm thick acrylic cruciform *via* commercially available cyanoacrylate adhesive and allowed to cure in air at room temperature for 48 hours.

**Raman and Photoluminescence Spectroscopy.** We took Raman and photoluminescence measurements using a Horiba Labram HR Evolution Raman System in the Stanford Nanofabrication Shared Facility. We used a green 532 nm laser with 1% incident laser power (0.5 mW) to avoid heating and material ablation. The laser spot size was less than 1 μm, and for maps, we used a spacing of at least 1.6 μm between each point to ensure that each measured spot on the sample was distinct. Peak fitting details can be found in Supporting Information Section 3.

**Electrical Measurements.** We used an in-air probe station with a Keithley 4200 semiconductor parameter analyzer to conduct our electrical measurements. All electrical measurements under strain were taken with the PEN sample attached to the acrylic cross to ensure that the sample attachment process did not impact the measurements. The measurements were performed in the dark at room temperature. Prior to measurement, the $AlO_x$ encapsulation layer was patterned and wet etched with Megaposit MF-26A developer for 10 mins to access the contact pads. Forward and backward voltage sweeps were taken for every measurement; a discussion of hysteresis can be found in Supporting Information Section 5.

**Density Functional Theory Calculations.** Density functional theory (DFT) calculations were performed using Quantum ESPRESSO[41] version 7.1. We use fully relativistic norm-conserving Vanderbilt pseudopotentials[42,43] and include the effect of spin-orbit coupling, which has been shown to significantly affect the band structure of $WS_2$.[31] All calculations were performed on 31 × 31 × 1 $k$-point grids with kinetic energy cutoffs of 70 Ry, and charge density and potential cutoffs of 560 Ry. Additional details can be found in Supporting Information Section 8.

## ASSOCIATED CONTENT

**Supporting Information**

The following file is available free of charge on the ACS Publications website. Device fabrication process, effect of encapsulation on $WS_2$ devices, Raman and photoluminescence spectroscopy measurement parameters and peak fitting, threshold voltage estimates, current-voltage characteristics of devices under strain, $WS_2$ devices at high strain and after strain relaxation, strain fields induced by the biaxial strain tool, density functional theory calculations of full $WS_2$ band structure, and channel length dependence of strain-induced mobility improvements.

## AUTHOR INFORMATION

**Corresponding Author**

*E-mail: epop@stanford.edu

**ORCID**

Jerry A. Yang: 0000-0001-5521-8523

Robert K. A. Bennett: 0000-0001-7427-8724

Zhepeng Zhang: 0000-0002-9870-0720

Andrew J. Mannix: 0000-0003-4788-1506

Eric Pop: 0000-0003-0436-8534

**Author Contributions**

The manuscript was written through contributions of all authors. All authors have given approval to the final version of the manuscript.

**Notes**

The authors declare no competing financial interest.

**ACKNOWLEDGMENTS**

This work was completed in part at the Stanford Nanofabrication and Stanford Nano Shared Facilities, which receive funding from the National Science Foundation (NSF) as part of the National Nanotechnology Coordinated Infrastructure (NNCI) Award ECCS-1542152. The work was also supported in part by the Stanford SystemX Alliance and by Intel Corporation. J.A.Y. acknowledges support from the NSF Graduate Research Fellowship. R.K.A.B acknowledges support from the Stanford Graduate Fellowship and NSERC PGS-D program. Z.Z. and A.J.M. acknowledge support from the US Department of Energy for the development of $WS_2$ synthesis, under award DE-SC0021984. K.J.T. acknowledges support from the Stanford Summer Undergraduate Research Program. We would like to thank Antonios Michail, Isha Datye, Marc Jaikissoon, Crystal Nattoo, Alwin Daus, and Ajay Sood for their valuable discussions.

**Land Acknowledgement.** This work was completed at Stanford University, on the ancestral land of the Muwekma Ohlone Tribe, which continues to be of great importance to the Ohlone people. We acknowledge and honor our institution's relationship to Indigenous peoples and their continual care for the land we use to conduct scientific research.[44]

## Table of Contents Figure

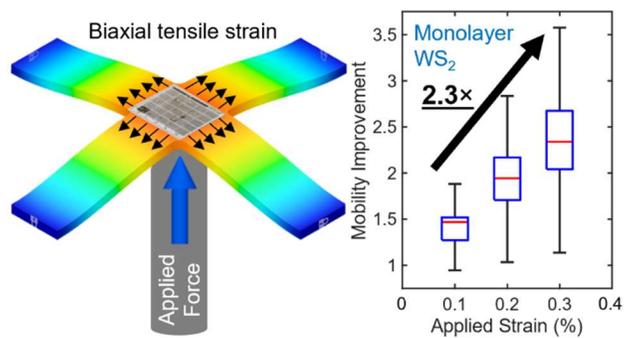

<p style="text-align:center;"><u>**Supporting Information**</u></p>

# Biaxial Tensile Strain Enhances Electron Mobility of Monolayer Transition Metal Dichalcogenides

Jerry A. Yang,[1] Robert K. A. Bennett,[1] Lauren Hoang,[1] Zhepeng Zhang,[2] Kamila J. Thompson,[1] Andrew J. Mannix[2,3], and Eric Pop[1,2,4,*]

[1]*Department of Electrical Engineering, Stanford University, Stanford, CA 94305, USA*
[2]*Department of Materials Science & Engineering, Stanford University, Stanford, CA 94305, USA*
[3]*Stanford Institute for Materials and Energy Sciences, SLAC National Accelerator Laboratory, Menlo Park, CA 94305, USA*
[4]*Precourt Institute for Energy, Stanford University, Stanford, CA 94305, USA*

*Corresponding Author: epop@stanford.edu

## 1. TMD device fabrication process

**Figure S1** describes and illustrates the WS$_2$ transistor fabrication process on polyethylene naphthalate (PEN) substrates of 2×2 cm$^2$. We first pattern the back-gate onto 125-μm thick polyethylene naphthalate (PEN) substrates with optical lithography, depositing Ti/Au (5 nm/40 nm) using electron-beam (e-beam) evaporation, and defining the back-gate by lift-off. We then deposit ~16 nm alumina (Al$_2$O$_3$) gate dielectric with atomic layer deposition (ALD) in a Savannah S200 from Cambridge Nanotech using trimethyl aluminum (TMA) and DI water as precursors for 200 cycles. The ALD was completed at 130 °C for ~3 hours to remain within the thermal budget of PEN.

We grow monolayer WS$_2$ on 50 mm diameter *c*-plane sapphire wafers, cleaved into quarters, by chemical vapor deposition (CVD) at 775 °C for 6 hours using diethyl sulfide (DES) and ammonium metatungstate (AMT) precursors, with potassium hydroxide growth promoter and Ar/H$_2$ carrier gases. Flow rates during growth are 0.05 sccm DES, 1600 sccm Ar, and 1 sccm H$_2$. The full details of this growth process will be discussed in a forthcoming paper. In addition, we grow monolayer MoS$_2$ by CVD on 2×2 cm$^2$ chips of 90 nm thermal SiO$_2$/Si wafers, using solid MoO$_3$ and S precursors.[1] The wafers are treated with an hexamethyldisilazane (HMDS) vapor prime, and drops of perylene-3,4,9,10-tetracarboxylic acid tetrapotassium acid salt (PTAS) are placed around the edges of the chips as a growth promoter. The samples are placed in a 50 mm diameter tube furnace with the precursors, and the growth is completed at 750 °C for 5 min at 800 Torr. 30 sccm Ar is used as the carrier gas for the growth.

After growth, we transfer the materials (either WS$_2$ or MoS$_2$) onto the samples with an approach similar to Vaziri *et al.*[2] We next define the rectangular channel using an O$_2$ plasma etch with a Samco PC3000. The etch conditions are 10 sccm O$_2$ at 75 mTorr at 100 W for 2 min. We then pattern and e-beam evaporate Ni/Au (5 nm/40 nm) source and drain contacts (Ni being in contact with the TMD). The channel definition is done before the contact deposition to anchor the WS$_2$ to the substrate during the strain measurements, as Ni has high adhesion energy to Al$_2$O$_3$ compared to Au alone.[3] Lastly, we encapsulate the WS$_2$ transistors with an AlO$_x$ (10 nm) layer similar to McClellan et al.[4] at 130 °C, which anneals and passivates the chiplet.

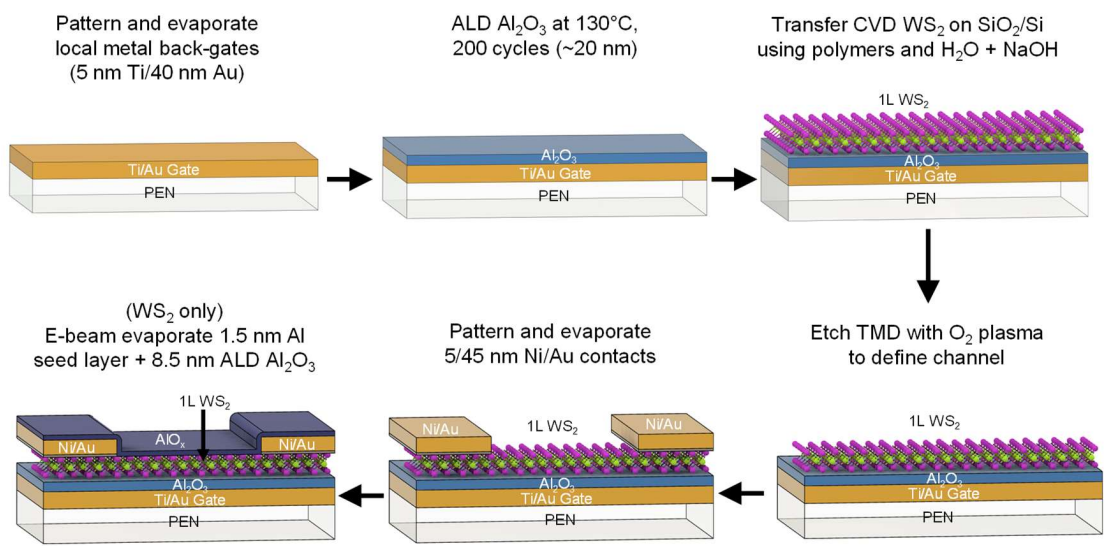

**Figure S1.** Summary of fabrication steps for WS$_2$ transistors on bendable PEN substrates. The fabrication of MoS2 transistors on PEN is identical, except for the final AlO$_x$ encapsulation.

We acquired a 3 mm thick polymethylmethacrylate (PMMA) cross custom-machined with computer numerical control (CNC) milling and annealed the cross for 30 min at 80 °C in an oven. Compared to laser cutting, CNC milling and annealing produced PMMA crosses that could tolerate greater than 1% strain without cracking. We attached the sample to the cross with commercial cyanoacrylate (Loctite), as shown in **Figure S2a**, and mounted the cross to the biaxial strain tool (**Figure S2b**).

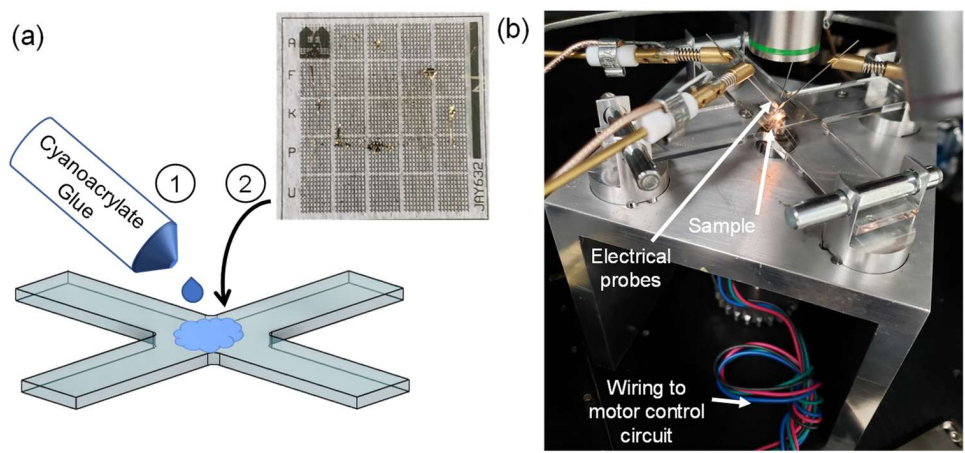

**Figure S2.** Sample attachment and mounting. (a) Sample attachment process with image of the PEN chiplet of 2×2 cm$^2$ size. Each chiplet has 5×5 (= 24 reticles plus a cat logo in the top-left corner) with 9×9 (=81) transistors each. (b) Biaxial strain tool with mounted sample on the cross under the microscope in the electrical probe station. The sample, electrical probes, and wiring to the motor control circuit are labeled. The acrylic cross is 3 mm thick, and its two arms are 115 mm long, from tip-to-tip.

3## 2. Effect of sub-stoichiometric AlO$_x$ encapsulation on WS$_2$ devices

**Figure S3** shows the $I_D$-$V_{GS}$ curves measured in air for 100 devices on the chiplet before and after AlO$_x$ encapsulation. Prior to encapsulation, many devices exhibited low current and significant hysteresis. This occurs because oxygen adsorption removes electrons from the WS$_2$ channel, shifting the $V_T$ positively, while charge trapping in the adsorbed molecules leads to large hysteresis in air ambient.[5,6] After encapsulation, the devices showed significant qualitative improvement. It appears that the encapsulation may passivate, electron dope,[4] as well as impart residual stress[7] into the WS$_2$ channel, improving the device performance and reducing device variability.

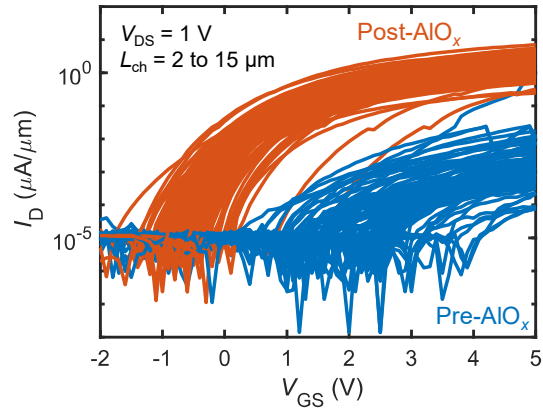

**Figure S3.** Measured $I_D$-$V_{GS}$ curves for 100 devices showing forward sweeps of the devices before (blue) and after (red) AlO$_x$ encapsulation at $V_{DS}$ = 1 V for $L_{ch}$ = 2 to 15 µm.

## 3. Raman and photoluminescence spectroscopy of WS$_2$ devices with strain

### 3.1 Peak fitting procedures

To track the appropriate peaks in the Raman and photoluminescence (PL) spectra of the WS$_2$ device channels, we fit the Raman and PL peaks using an iterative least-squares method implemented in MATLAB.[8] All Raman spectra were taken between ~100 cm$^{-1}$ and ~550 cm$^{-1}$, and all PL spectra were taken between 1.7 eV and 2.1 eV. The baselines of the spectra were subtracted prior to fitting. For the Raman spectra of unencapsulated devices, four peaks were used to fit the spectral region between 300 cm$^{-1}$ and 400 cm$^{-1}$, and one peak was used for the A$_1$' phonon mode between 400 cm$^{-1}$ and 440 cm$^{-1}$. For the Raman spectra of encapsulated devices, only three peaks were used to fit the spectral region between 300 cm$^{-1}$ and 400 cm$^{-1}$ because the four-peak fit resulted in greater residual error. For the PL spectra of all devices, two peaks were used to fit the PL spectrum between 1.7 eV and 2.1 eV. We used an equally-weighted Gaussian-Lorentzian line shape for all peaks. The black lines in **Figure 2a and 2c** indicate the peaks used to fit the spectra.

### 3.2 Strain-dependent measurements of unencapsulated WS$_2$ devices

We measure the strain-induced peak shift rates for unencapsulated WS$_2$ devices to confirm that the biaxial strain tool imparts tensile strain to the TMD samples. **Figure S5** shows the Raman and PL spectra, as well as the extracted strain rates, for a single device on a second WS$_2$ sample prepared during the same fabrication run. Error bars for the extracted strain rates are calculated with a 95% confidence interval. While the data show large error bars because the E' and 2LA(M) modes are broad and only ~4 cm$^{-1}$ apart and are difficult to distinguish with certainty using least-squares methods,[9] our estimated strain rates for the device match previous literature using a similar biaxial strain mechanism.[10] Comparing **Figure S4** (unencapsulated) with **Figure 2** in the main text (encapsulated with AlO$_x$) we note some effects of the AlO$_x$ encapsulation step, which removes surface oxygen and water adsorbates, inducing partial electron doping in WS$_2$. The encapsulation step also leads to more consistent optical measurements (**Figure 2**) during the long biaxial strain measurement sessions (several hours).

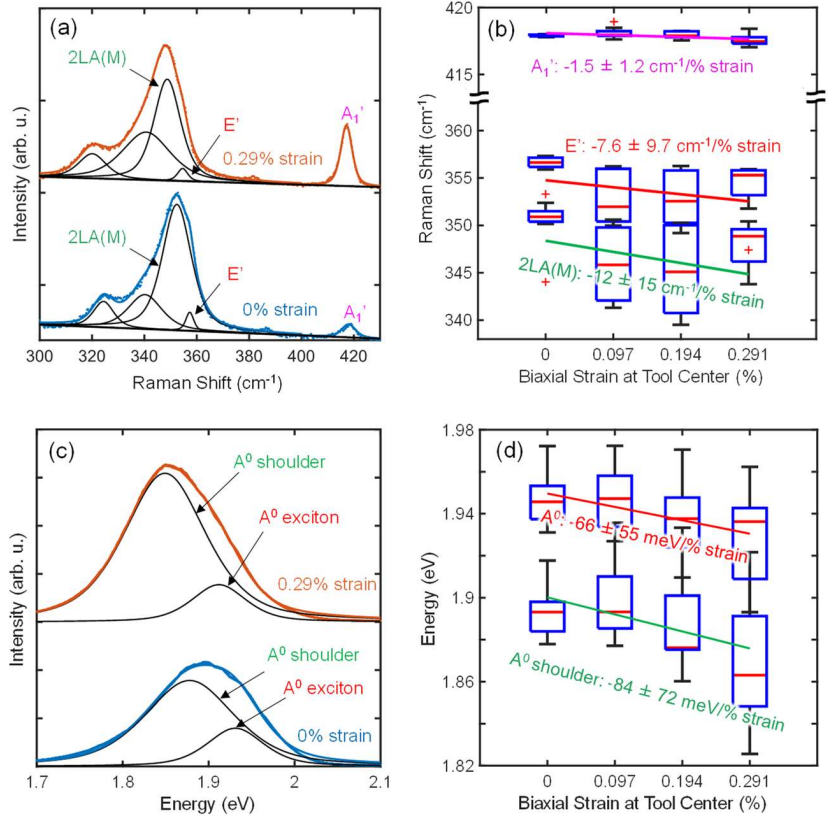

**Figure S4.** Strain-dependent Raman and PL measurements of an unencapsulated WS$_2$ device. (a) Raman spectra of a device with similar geometry as **Figure 1b** in the main text without biaxial strain (blue) and with biaxial strain applied (orange). (b) Extracted phonon shift rates of the A$_1$' (pink), E' (red), and 2LA(M) (green) Raman peaks of the unencapsulated WS$_2$ as a function of biaxial strain applied, with box plots indicating the median (red line), first, and third quartiles (blue box) across nine measured locations in the device channel. Red plus '+' symbols indicate outliers. The large error bars for the E' and 2LA(M) peaks are due to their proximity to each other (~4 cm$^{-1}$) and the significantly higher intensity of the 2LA(M) peak compared to the E' peak, making them difficult to distinguish using iterative computational peak fitting methods.[9] (c) PL spectra of the device at the same locations in the channel, under the same conditions and color labeling as in (a). (d) Extracted exciton shift rates of the A$^0$ exciton (red) and its left shoulder peak (green), with similar notation and same number of points as in (b).

## 4. Threshold voltage ($V_\text{T}$), on-current ($I_\text{on}$), and mobility ($\mu_\text{FE}$) estimates

For all data reported in the main text, we estimate the threshold voltage ($V_\text{T}$) of our devices using the constant-current method[11] with threshold current $I_\text{T}$ = 10 nA/μm based on the International Roadmap for Devices and Systems (IRDS) 2022 threshold for high-performance devices.[12] When this threshold current was captured between two measured drain current ($I_D$) values, we used logarithmic interpolation to estimate the $V_{GS} = V_T$ at which the device would have $I_D = 10$ nA/μm.

We set the gate overdrive $V_{GS} - V_T = 5$ V to the largest integer voltage such that $I_{on}$ and $\mu_{FE}$ could be extracted for all devices. When $V_{GS} = V_T + 5$ V falls between two instrument-measured values we use linear interpolation to estimate the $I_{on} = I_D$ at that point. At the same $V_{GS}$ we also estimate $\mu_{FE}$ as:

$$\mu_{FE} = \frac{g_m L_{ch}}{W C_{ox} V_{DS}} = \left(\frac{\partial I_D}{\partial V_{GS}}\right)\frac{L_{ch}}{W C_{ox} V_{DS}}.$$

Although we use $V_T$ from the constant-current[11] method throughout the main text, we also compare it with $V_T$ from linear extrapolation[11] and Y-function[13] methods for both WS$_2$ and MoS$_2$ in **Figure S5**.

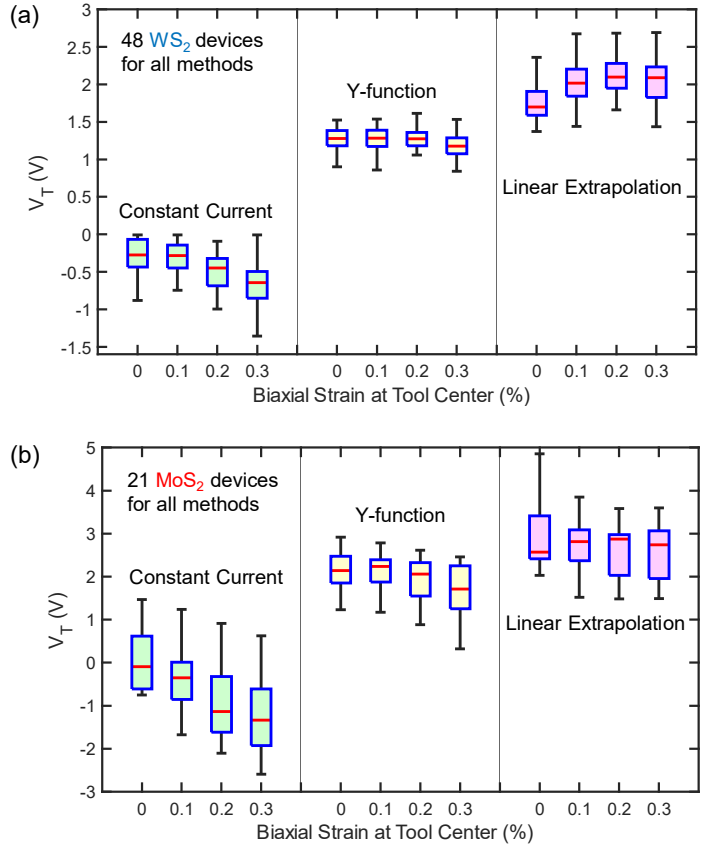

**Figure S5.** Comparison of $V_T$ extraction methods for constant current (left), Y-function (middle), and linear extrapolation (right) for the (a) 48 WS$_2$ devices and (b) 21 MoS$_2$ devices. $V_T$ is reported as a function of applied biaxial strain, from 0% (unstrained) to 0.3%.

In **Figure S5**, we note that threshold voltages are generally around zero for our devices, but there are several discrepancies between $V_T$ extracted from various methods. These have been previously observed in the TMD device literature[1,14] and they are due to at least two causes: 1) the constant-current $V_T$ is somewhat arbitrary, and a larger $V_T$ can be extracted if this is done at a larger current, *e.g.* $I_D = 100$ nA/μm, and 2) the measured $I_D$-$V_{GS}$ curves (on linear axes) often display a "round knee" at lower $V_{GS}$, before reaching the true linear region at higher $V_{GS}$. This is partly due to the contacts "turning on" at different $V_{GS}$ from the channel[14–16] and partly due to more trap-assisted (hopping) transport at lower $V_{GS}$, closer to threshold.[17,18] Hence, we report all our data (for $I_{on}$ and $\mu_{FE}$) at "as high a $V_{GS}$ as possible" everywhere else in our study, to be in the linear region of our devices.

We recall that in the main text (**Figures 3c,d** and **4c,d**), we reported all our findings at $V_{GS} - V_T = 5$ V, which was the highest $V_{GS}$ in the linear region for a substantial number of devices (48 for WS$_2$ and 21 for MoS$_2$). For comparison, in **Figure S6**, we also report the $I_{on}$ and $\mu_{FE}$ at a lower $V_{GS} - V_T = 4$ V, which now includes 60 WS$_2$ and 20 MoS$_2$ devices. Overall, the improvements with applied biaxial strain are slightly lower than those at $V_{GS} - V_T = 5$ V reported in the main text. Compare, for example, **Figure S6a,b** below with **Figure 3c,d** in the main text. This suggests that strain-induced device improvements may be slightly better at higher gate overdrive, potentially due to better contact gating (and reduced contact resistance) at higher $V_{GS}$. However, there is sufficient uncertainty in the average $I_{on}$ improvement (*e.g.* 1.8 ± 0.5 times at 0.3% strain, see **Figure 3c**) that its root cause remains unclear, and it will be the subject of future studies.

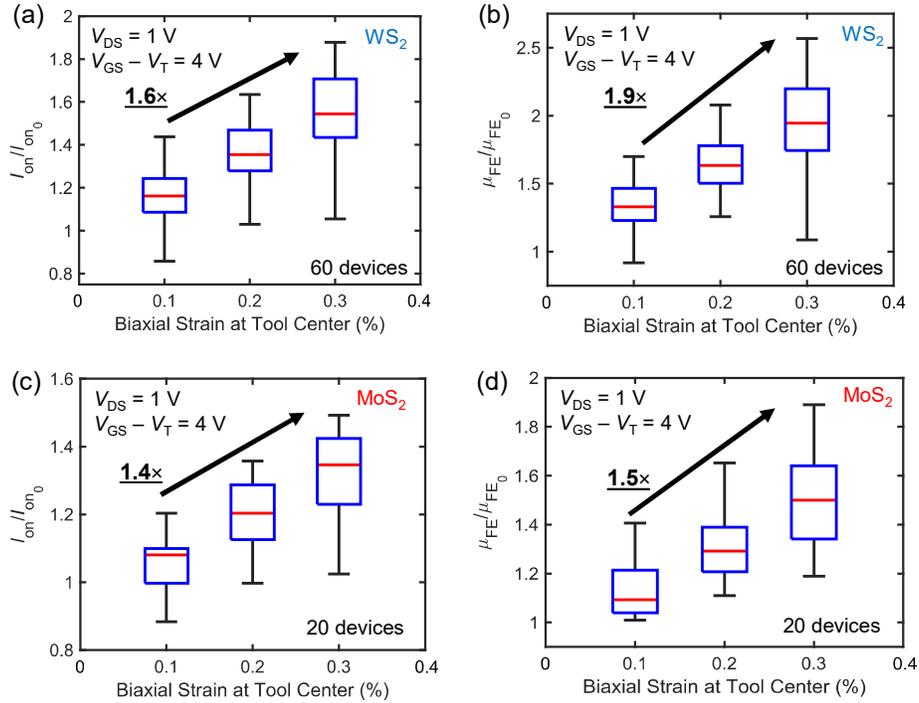

**Figure S6.** Box plots of normalized $I_{on}$ and $\mu_{FE}$ improvement as a function of applied biaxial tensile strain similar to **Figures 3c,d** and **4c,d** in the main text, extracted at $V_{DS} = 1$ V and $V_{GS} - V_T = 4$ V. (a-b) display relative $I_{on}$ and $\mu_{FE}$ changes across 60 WS$_2$ devices and (c-d) are across 20 MoS$_2$ devices. The $I_{on}$ and $\mu_{FE}$ improve by 1.6 ± 0.14 and 1.9 ± 0.19 respectively for WS$_2$, and 1.4 ± 0.25 and 1.5 ± 0.40 respectively for MoS$_2$ devices. For comparison, those in the main text figures are at a larger gate overdrive, $V_{GS} - V_T = 5$ V.

## 5. Current-voltage characteristics of TMD transistors with strain

We measure our devices electrically with a Keithley 4200-SCS using Keithley Interactive Test Environment (KITE) software. **Figure S7** displays the same $I_D$-$V_{GS}$ curves for the devices in **Figures 3a** and **4a** of the main text, but here includes the reverse sweep (dashed lines) and gate current (dotted lines) with the forward sweep (solid lines). The mismatched forward and backward sweeps point to hysteresis in the devices, likely arising from electrically active interfacial traps between the TMD and Al$_2$O$_3$ due to adsorbates from the transfer process.[19] The gate current ($I_G$) remains at or below the tool measurement threshold (~100 pA) throughout all four WS$_2$ measurements, indicating no substantial leakage current through the gate dielectric. The Keithley 4200-SCS settings were later adjusted to measure lower gate currents more accurately for the MoS$_2$ transistors.

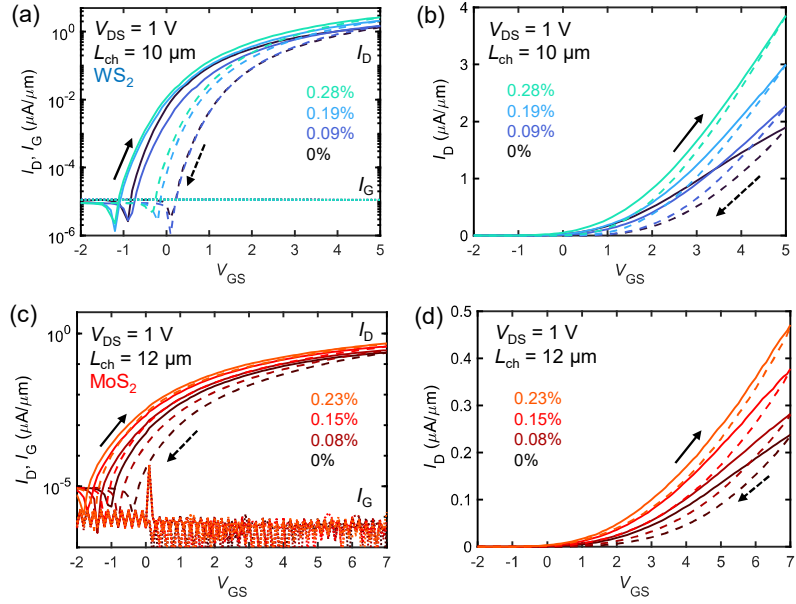

**Figure S7.** (a-b) Measured $I_D$-$V_{GS}$ curves showing forward (solid) and backward (dashed lines) sweeps at different strain levels for the same WS$_2$ device (a-b) shown in **Figure 3a** of the main text, with $W$ = 20 μm, $L_{ch}$ = 10 μm, and $V_{DS}$ = 1 V, on (a) logarithmic scale and (b) linear scale. The gate current ($I_G$) is below the tool measurement range setting (~100 pA). (c-d) Measured $I_D$-$V_{GS}$ for the MoS$_2$ device from **Figure 4a** of the main text, with $W$ = 20 μm, $L_{ch}$ = 12 μm, and $V_{DS}$ = 1 V. The gate current, $I_G$, is also normalized by the channel width, like the drain current.

**Figure S8** plots the hysteresis (defined as the difference in $V_T$ between the reverse and forward sweeps divided by the voltage sweep range) vs. strain for the 48 WS$_2$ devices included in **Figure 3c-d** and the 21 MoS$_2$ devices in **Figure 4c-d**. We observe negligible dependence of hysteresis on strain, within the range applied here, for our devices.

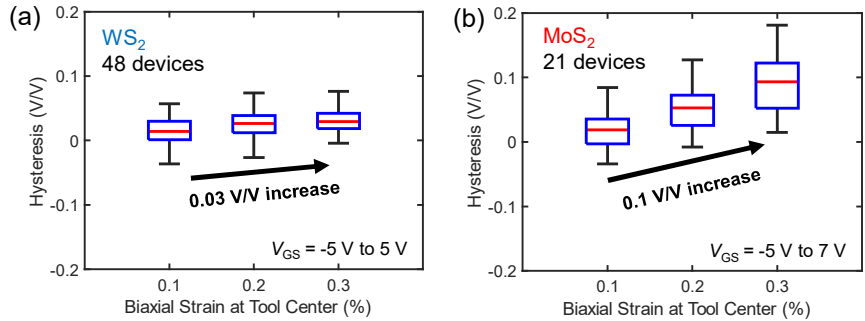

**Figure S8.** Box plots showing extracted hysteresis values for (a) 48 WS$_2$ devices, and (b) 21 MoS$_2$ devices as a function of biaxial tensile strain. The hysteresis is computed as the ratio of ($V_{T,reverse} - V_{T,forward}$) divided by the voltage sweep range (here, 10 V for WS$_2$ and 12 V for MoS$_2$, although only the ranges of interest are shown in **Figure S7**).

4## 6. WS$_2$ devices at high biaxial strain and after strain relaxation

We measured the WS$_2$ devices at 0.4% applied biaxial strain and after strain relaxation. (Similar measurements for uniaxially strained and relaxed MoS$_2$ devices on flexible substrates were carried out by Datye et al.[20]) The relaxed measurement was completed several days after the initial strain measurement to ensure that residual creep strain in the acrylic cross had fully relaxed.

**Figure S9** shows one device at 0.4% applied biaxial strain and after strain relaxation. The device exhibited poorer $I_D$ at 0.4% strain and gate leakage after relaxation. The lower $I_D$ suggests poorer contact adhesion at high strain, and the increased gate leakage ($I_G$) indicates that the Al$_2$O$_3$ gate dielectric likely suffered mechanical defects under the shear strains imparted by the tool.

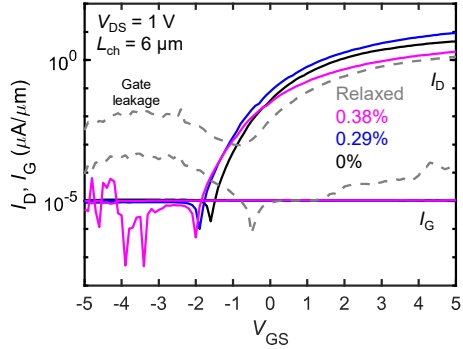

**Figure S9.** Measured drain ($I_D$) and gate ($I_G$) current vs. $V_{GS}$ at different strain levels for a device with $W$ = 20 μm, $L_{ch}$ = 6 μm, and $V_{DS}$ = 1 V. Black, blue, magenta, and gray correspond to 0%, 0.29%, 0.38% strain (solid lines) and after strain relaxation (dashed lines), respectively.

**Figure S10** shows the box plots of $I_{on}$ and $\mu_{FE}$ for the 15 working devices that survived the higher-strain (up to 0.4%) and return to a relaxed state. While up to 0.3% strain the trend is clear, at 0.4% strain over 75% of the 48 measured devices either exhibited shorted gate-to-contact behavior or showed significant current degradation. The shorted gate indicates that the strain likely caused microcracks in the Al$_2$O$_3$ that further broke down under electrical stress.[21] In addition, the current degradation occurred likely due to slippage at the WS$_2$-contact interfaces.[20] After relaxation, the surviving devices (on average) return to their original (unstrained) $I_{on}$ and $\mu_{FE}$, albeit with a broader distribution due to the devices that exhibited partial failure (note the lower bar of the box plot near the bottom axis).

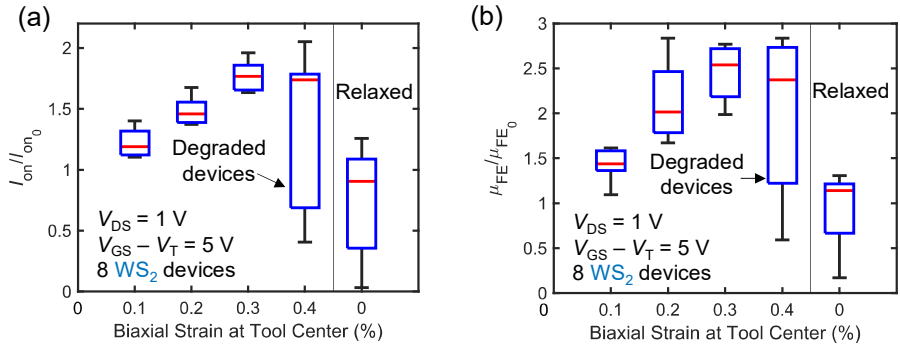

**Figure S10.** Box plots of the normalized (a) $I_{on}$ and (b) $\mu_{FE}$ for the remaining 8 devices whose $V_T$ could be extracted after the strain measurements. Three of the devices showed significant degradation at 0.4% strain, creating the large spread of the box plot, while the other five largely returned to their original $I_{on}$ and $\mu_{FE}$ after relaxation.

## 7. Strain fields induced by biaxial strain tool

Because the deflection occurs at the center of the cross, the strain field in the sample for each strain level is approximately radial with the maximum strain at the center of the tool.[22] **Figure S11** shows the spatial distribution of measured devices for each strain level on the $WS_2$ sample, with the colors indicating the improvements in mobility, relative to the unstrained case. Devices near the center of the tool appear to have larger improvements from 0.1%-0.3% strain and degrade at 0.4% strain.[22]

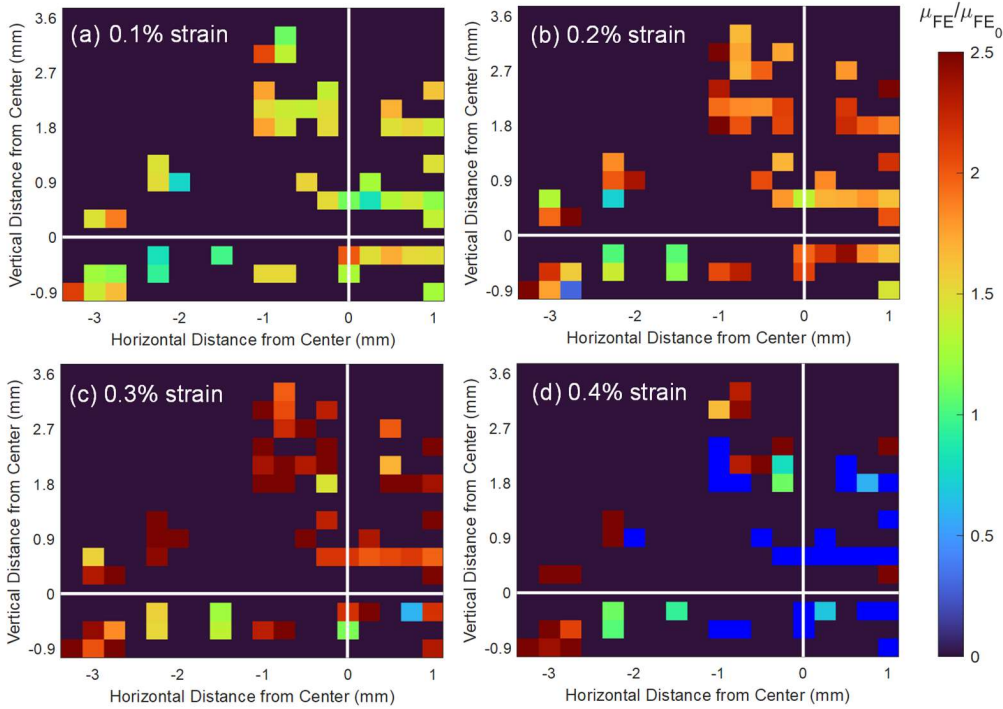

**Figure S11.** Heat maps showing the spatial distribution of the devices across the sample relative to the center of the strain tool and their corresponding $\mu_{FE}$ (normalized with respect to the unstrained devices). (a) At 0.1% applied biaxial tensile strain, (b) at 0.2%, (c) at 0.3%, (d) at 0.4%. On average, mobility improves from 0.1% to 0.3% applied strain. At 0.4%, devices near the center of the cross begin to degrade, as seen by the sharp drop of mobility (blue tiles). Dark blue tiles in panel (d) represent devices whose mobilities could not be extracted. Only a subset of the functioning fabricated devices (48 colored "tiles" in each panel) were measured here.

## 8. Density functional theory (DFT) simulations of strained $WS_2$ band structure

We employ DFT calculations with spin-orbit coupling to understand the causes of mobility enhancement in our biaxially-strained devices. We calculate the band structure for $WS_2$ under no strain and from 0.1% to 1% uniaxial and biaxial strain. **Figure S12** shows the calculated band structure relative to the conduction band edge, and the reduction of the energy band gap ($E_G$) with tensile strain. We observe a ~40 meV shift in $V_T$ with 0.3% biaxial tensile strain applied in this study, which is small compared to our gate voltage overdrive used for mobility extraction. We also note that the calculated reduction of $E_G$ by -128 ± 0.41 meV/% biaxial tensile strain closely matches the measured $A^0$ exciton shift rate in the PL spectra of **Figure 2(c,d)**.

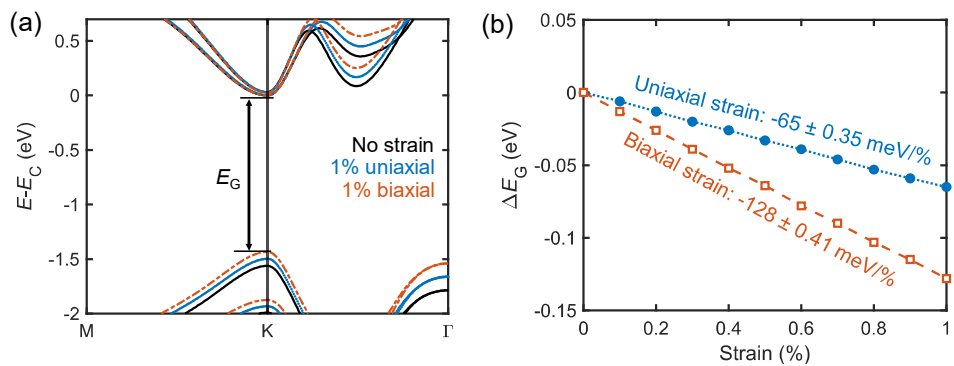

**Figure S12.** DFT calculations with spin-orbit coupling. (a) Calculated band structure of $WS_2$ with no strain (black solid line), 1% uniaxial tensile strain (blue solid line), and 1% biaxial tensile strain (orange dotted-dashed line) relative to the conduction band edge. (b) Calculated energy band gap reduction ($\Delta E_G$) for uniaxial (blue dotted line) and biaxial (orange dashed line) tensile strain from 0% to 1%. Markers indicate calculated points.

## 9. Channel length dependence of strain-induced mobility improvements

To confirm that the strain-induced improvement of mobility is not a contact effect, we plot the mobility improvement against channel length in **Figure S13**. We observe that the device variability within each channel length is greater than the variation across channel lengths, indicating that the local channel material quality dominates over contact effects in our data.

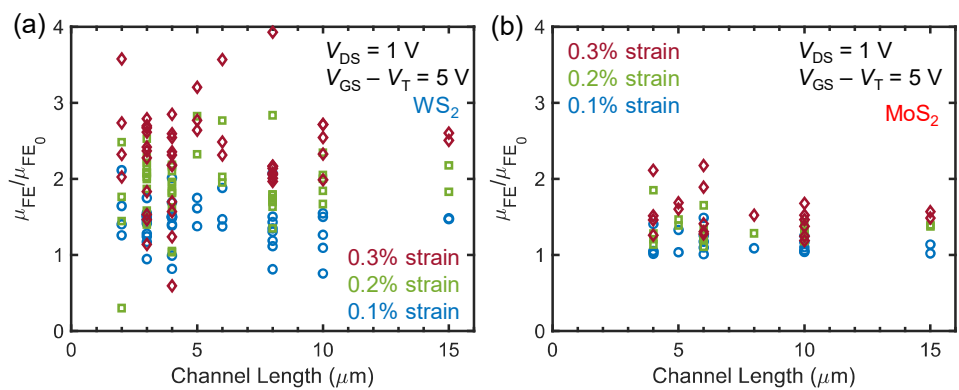

**Figure S13.** Channel length dependence of strain-induced mobility improvement in (a) $WS_2$ and (b) $MoS_2$ transistors. Each symbol represents one device measured at one strain point, with blue circles representing 0.1% biaxial tensile strain, green squares representing 0.2% strain, and red diamonds representing 0.3% strain. Mobilities are estimated at a gate voltage overdrive of 5 V with constant-current $V_T$.

## 10. Supplementary References

11(3) Li, H.-T.; Chen, L.-F.; Yuan, X.; Zhang, W.-Q.; Smith, J. R.; Evans, A. G. Interfacial Stoichiometry and Adhesion at Metal/α-Al2O3 Interfaces. *J. Am. Ceram. Soc.* **2011**, *94* (s1), s154–s159. https://doi.org/10.1111/j.1551-2916.2011.04405.x.

(4) McClellan, C. J.; Yalon, E.; Smithe, K. K. H.; Suryavanshi, S. V.; Pop, E. High Current Density in Monolayer MoS$_2$ Doped by AlO$_x$. *ACS Nano* **2021**, *15* (1), 1587–1596. https://doi.org/10.1021/acsnano.0c09078.

(5) Liang, Q.; Gou, J.; Arramel; Zhang, Q.; Zhang, W.; Wee, A. T. S. Oxygen-Induced Controllable p-Type Doping in 2D Semiconductor Transition Metal Dichalcogenides. *Nano Res.* **2020**, *13* (12), 3439–3444. https://doi.org/10.1007/s12274-020-3038-8.

(6) Urban, F.; Giubileo, F.; Grillo, A.; Iemmo, L.; Luongo, G.; Passacantando, M.; Foller, T.; Madauß, L.; Pollmann, E.; Geller, M. P.; Oing, D.; Schleberger, M.; Bartolomeo, A. D. Gas Dependent Hysteresis in MoS$_2$ Field Effect Transistors. *2D Mater.* **2019**, *6* (4), 045049. https://doi.org/10.1088/2053-1583/ab4020.

(7) Ylivaara, O. M. E.; Liu, X.; Kilpi, L.; Lyytinen, J.; Schneider, D.; Laitinen, M.; Julin, J.; Ali, S.; Sintonen, S.; Berdova, M.; Haimi, E.; Sajavaara, T.; Ronkainen, H.; Lipsanen, H.; Koskinen, J.; Hannula, S.-P.; Puurunen, R. L. Aluminum Oxide from Trimethylaluminum and Water by Atomic Layer Deposition: The Temperature Dependence of Residual Stress, Elastic Modulus, Hardness and Adhesion. *Thin Solid Films* **2014**, *552*, 124–135. https://doi.org/10.1016/j.tsf.2013.11.112.

(8) O'Haver, T. Interactive Peak Fitter, 2021. https://terpconnect.umd.edu/~toh/spectrum/InteractivePeakFitter.htm.

(9) McCreary, A.; Berkdemir, A.; Wang, J.; Nguyen, M. A.; Elías, A. L.; Perea-López, N.; Fujisawa, K.; Kabius, B.; Carozo, V.; Cullen, D. A.; Mallouk, T. E.; Zhu, J.; Terrones, M. Distinct Photoluminescence and Raman Spectroscopy Signatures for Identifying Highly Crystalline WS$_2$ Monolayers Produced by Different Growth Methods. *J. Mater. Res.* **2016**, *31* (7), 931–944. https://doi.org/10.1557/jmr.2016.47.

(10) Michail, A.; Anestopoulos, D.; Delikoukos, N.; Grammatikopoulos, S.; Tsirkas, S. A.; Lathiotakis, N. N.; Frank, O.; Filintoglou, K.; Parthenios, J.; Papagelis, K. Tuning the Photoluminescence and Raman Response of Single-Layer WS$_2$ Crystals Using Biaxial Strain. *J. Phys. Chem. C* **2023**, *127* (7), 3506–3515. https://doi.org/10.1021/acs.jpcc.2c06933.

(11) Ortiz-Conde, A.; García Sánchez, F. J.; Liou, J. J.; Cerdeira, A.; Estrada, M.; Yue, Y. A Review of Recent MOSFET Threshold Voltage Extraction Methods. *Microelectron. Reliab.* **2002**, *42* (4), 583–596. https://doi.org/10.1016/S0026-2714(02)00027-6.

(12) *International Roadmap for Devices and Systems 2022*; Institute of Electrical and Electronics Engineers, 2022. https://irds.ieee.org/editions/2022 (accessed 2023-07-07).

(13) Pacheco-Sanchez, A.; Jiménez, D. Accuracy of Y-Function Methods for Parameters Extraction of Two-Dimensional FETs across Different Technologies. *Electron. Lett.* **2020**, *56* (18), 942–945. https://doi.org/10.1049/el.2020.1502.

(14) Sebastian, A.; Pendurthi, R.; Choudhury, T. H.; Redwing, J. M.; Das, S. Benchmarking Monolayer MoS$_2$ and WS$_2$ Field-Effect Transistors. *Nat. Commun.* **2021**, *12* (1), 693. https://doi.org/10.1038/s41467-020-20732-w.

(15) Pang, C.-S.; Zhou, R.; Liu, X.; Wu, P.; Hung, T. Y. T.; Guo, S.; Zaghloul, M. E.; Krylyuk, S.; Davydov, A. V.; Appenzeller, J.; Chen, Z. Mobility Extraction in 2D Transition Metal Dichalcogenide Devices—Avoiding Contact Resistance Implicated Overestimation. *Small* **2021**, *17* (28), 2100940. https://doi.org/10.1002/smll.202100940.